\documentclass[%
 aip,
 apl,    
 amsmath,amssymb,
 reprint, 
]{revtex4-1}

\usepackage{graphicx} 
\usepackage{dcolumn} 
\usepackage{bm} 
\usepackage[utf8]{inputenc}
\usepackage[T1]{fontenc}
\usepackage{mathptmx}
\usepackage{etoolbox}
\usepackage{xcolor}
\usepackage{soul}
\usepackage{siunitx}
\usepackage{mathtools,amssymb}
\DeclareSIUnit\sq{\ensuremath{\Box}}
\DeclareSIUnit{\cps}{\text{cps}}
\DeclareSIUnit\sccm{sccm}

\usepackage[hidelinks]{hyperref}

\makeatletter
\def\@email#1#2{%
 \endgroup
 \patchcmd{\titleblock@produce}
  {\frontmatter@RRAPformat}
  {\frontmatter@RRAPformat{\produce@RRAP{*#1\href{mailto:#2}{#2}}}\frontmatter@RRAPformat}
  {}{}
}%
\makeatother
\begin{document}
\preprint{AIP/123-QED}

\title{Mid-infrared characterization of NbTiN superconducting nanowire single-photon detectors on silicon-on-insulator}

\author{Adan Azem}
\affiliation{Department of Electrical and Computer Engineering, University of British Columbia, Vancouver, B.C. V6T 1Z4, Canada} 
\affiliation {Stewart Blusson Quantum Matter Institute, University of British Columbia, Vancouver, B.C. V6T 1Z4, Canada}

\author{Dmitry V. Morozov}
\affiliation{James Watt School of Engineering, University of Glasgow, Glasgow, G12 8QQ, United Kingdom}

\author{Daniel Kuznesof}
\affiliation{James Watt School of Engineering, University of Glasgow, Glasgow, G12 8QQ, United Kingdom}

\author{Ciro Bruscino}
\affiliation{Dip. di Fisica, E. Pancini, Università degli Studi di Napoli Federico II, I-80125 Napoli, Italy}

\author{Robert H. Hadfield}
\affiliation{James Watt School of Engineering, University of Glasgow, Glasgow, G12 8QQ, United Kingdom}

\author{Lukas Chrostowski}
\affiliation{Department of Electrical and Computer Engineering, University of British Columbia, Vancouver, B.C. V6T 1Z4, Canada}
\affiliation{Stewart Blusson Quantum Matter Institute, University of British Columbia, Vancouver, B.C. V6T 1Z4, Canada}

\author{Jeff F. Young}
\affiliation {Stewart Blusson Quantum Matter Institute, University of British Columbia, Vancouver, B.C. V6T 1Z4, Canada}
\affiliation{Department of Physics and Astronomy, University of British Columbia, Vancouver, B.C. V6T 1Z1, Canada}

\date{\today}

\begin{abstract}
Superconducting nanowire single-photon detectors are widely used for detecting individual photons across various wavelengths from ultraviolet to near-infrared range. Recently, there has been increasing interest in enhancing their sensitivity to single photons in the mid-infrared spectrum, driven by applications in quantum communication, spectroscopy and astrophysics. Here, we present our efforts to expand the spectral detection capabilities of U-shaped $NbTiN$-based superconducting nanowire single-photon detectors, fabricated in a 2-wire configuration on a silicon-on-insulator substrate, into the mid-infrared range. We demonstrate saturated internal detection efficiency extending up to a wavelength of \SI{3.5}{\micro\meter} for a \SI{5}{\nano\meter} thick and \SI{50}{\nano\meter} wide $NbTiN$ nanowire with a dark count rate less than 10 counts per second at \SI{0.9}{\kelvin} and a rapid recovery time of \SI{4.3}{\nano\second}. The detectors are engineered for integration on waveguides in a silicon-on-insulator platform for compact, multi-channel device applications.
\end{abstract}

\maketitle
Superconducting nanowire single-photon detectors (SNSPDs) offer exceptional sensitivity across the ultraviolet to near-infrared (NIR) and short-wave infrared (SWIR) spectrum \cite{verevkin_detection_2002, marsili_detecting_2013, slichter_uv-sensitive_2017, morozov_superconducting_2021}, accompanied by unparalleled timing performance \cite{korzh_wsi_2018, korzh_demonstration_2020}, positioning them at the forefront of commercially available single-photon detection technologies \cite{you_superconducting_2020}. Years of effort have resulted in near-unity detection efficiency out to NIR wavelengths of $\sim$ \SI{1550}{\nano\meter} \cite{marsili_detecting_2013, chang_detecting_2021}.
Recent emerging applications in quantum-enhanced communication \cite{temporao_feasibility_2008}, computing \cite{yan_silicon_2021}, spectral sensing and
imaging \cite{chen_mid-infrared_2017, dello_russo_advances_2022,lau_superconducting_2023}, light detection and ranging (LiDAR) \cite{taylor_photon_2019, hadfield_single-photon_2023}, and astrophysics \cite{wollman_recent_2021} require similarly high-efficiency single photon sensitivity at longer wavelengths, particularly in the mid-infrared (MIR) range.

The overall {\it system} detection efficiency of SNSPDs is determined by a product of the i) internal detection efficiency (IDE), ii) the external coupling efficiency of light-to-be-detected onto the actual detector element, and iii) the absorption efficiency of the coupled light by the superconducting nanowire within the device \cite{ferrari_waveguide-integrated_2018}. Of these three, the IDE is the factor that is fundamentally reduced at longer wavelengths as the energy per photon decreases, resulting in less energy transfer to the nanowire \cite{lau_superconducting_2023}. Though relatively fragile, low-loss optical fibers exist in the MIR, and various electromagnetic cavity designs developed for fiber-coupled NIR detectors should port over to the MIR with relative ease \cite{shankar_mid-infrared_2011}.

The IDE is determined by a variety of material, environmental and geometric factors \cite{holzman_superconducting_2019}. One approach to improving MIR performance is to form the nanowires from superconducting material with low superconducting gap energy to compensate for the reduced energy per photon \cite{chen_mid-infrared_2021}. Among these materials, tungsten silicide ($WSi$) has emerged as a popular choice for MIR SNSPDs. Recent extensive engineering efforts have demonstrated its saturated IDE up to \SI{29}{\micro\meter} \cite{taylor_low-noise_2023}. However, due to its low superconducting gap, $WSi$ has a low critical temperature and small critical current, which imposes stringent requirements on the cooling system and readout circuitry. Alternatively, standard Nb-based superconductors can be engineered for enhanced MIR sensitivity. Niobium titanium nitride ($NbTiN$) is favoured for SNSPDs \cite{tanner_enhanced_2010} due to its relatively high critical temperature ($T_c$), low kinetic inductance ($L_k$) \cite{miki_superconducting_2009}, and ease of thin film deposition \cite{iosad_optimization_1999} and fabrication. Moreover, it offers tunability of thin film properties by controlling the deposition conditions \cite{pratap_optimization_2023}.

Regardless of the elemental composition of the superconductor, other ways to maximize the response at longer wavelengths include (1) reducing the operation temperature to increase the accessible bias current range \cite{goltsman_middle-infrared_2007}, (2) minimizing the cross-sectional area of the superconducting nanowire (i.e., reducing its thickness or width) \cite{marsili_efficient_2012, chang_efficient_2022, taylor_mid-infrared_2022}, and (3) decreasing the free carrier density by increasing the thin film normal state sheet resistance \cite{verma_single-photon_2021, colangelo_large-area_2022}. These approaches have been employed in MIR SNSPDs using several superconducting materials with meander-shaped or single-strip nanowires .

\begin{figure*} 
\includegraphics{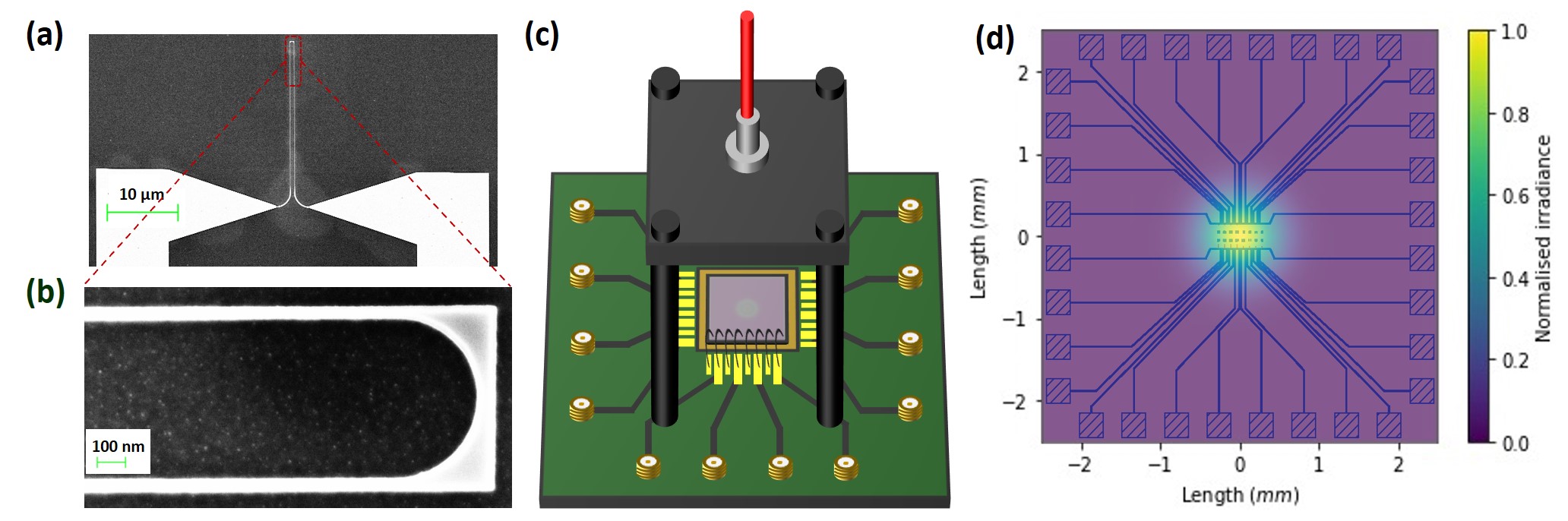}
\caption{(a) An SEM image of a single SNSPD in a 2-wire configuration, showing a U-shaped nanowire with tapered ends connected to electrical leads. (b) An SEM image of the far end of a U-shaped \SI{50}{\nano\meter} wide $NbTiN$ nanowire. (c) A schematic of the flood illumination measurement setup housed inside a cryostat combining a pulse tube cryocooler with a $^{4}$He sorption stage. The diagram shows the chip under test in the center (grey), a PCB (green), a copper plate under the chip (orange), a fiber port mount (black), a MIR fiber (red), and some electrical wire bonds (d) Normalized irradiance on the chip surface at a wavelength of \SI{2900}{\nano\meter}, calculated using the beam divergence model and fiber spec.}
\label{fig:chip_image}
\end{figure*}

The objective of the work reported here was to demonstrate near unity IDE beyond \SI{3}{\micro\meter} using a $NbTiN$ nanowire patterned on silicon-on-insulator (SOI) in a geometry compatible with future waveguide integration \cite{ferrari_waveguide-integrated_2018}. This work was originally inspired by a recent perspective paper \cite{yan_silicon_2021}, where we outlined a roadmap for developing long coherence time spin qubits in a silicon photonic circuit-based platform using singly ionized selenium, which introduces an optical transition at \SI{2.9}{\micro\meter} into the silicon band structure \cite{morse_photonic_2017}. Employing an SOI platform with a \SI{500}{\nano\meter} thick silicon photonic layer enables low-loss single-mode operation at this wavelength.  More generally, waveguide-integrated SNSPD would allow the opportunity to develop a variety of complex photonic-based quantum information processing systems \cite{moody_2022_2022}, offering reduced detection noise, dense integration with other photonic components, improved photon collection, and controlled photon propagation. Compared to meandered SNSPDs, waveguide-integrated SNSPDs enable the utilization of shorter nanowires, which are typically more uniform with a lower probability of constrictions. This results in relaxed fabrication requirements and faster timing response while preserving high absorption efficiency.

Here, we report expanding the spectral detection range of free-space coupled U-shaped $NbTiN$-based SNSPDs in a 2-wire configuration (not a transmission line) fabricated on a \SI{500}{\nano\meter} SOI substrate. By utilizing this standard design for waveguide-integrated SNSPDs and patterning the detectors on an SOI photonic platform, our devices are optimally suited for waveguide integration. The SNSPDs exhibit saturated IDE from \SI{1550}{\nano\meter} up to \SI{2900}{\nano\meter} and near-unity IDE at \SI{3502}{\nano\meter}. Additionally, a dark count rate (DCR) of less than 10 counts per second (\SI{}{\cps}) is demonstrated at a temperature of \SI{0.9}{\kelvin}. The results represent a significant step towards the development of integrated MIR quantum photonic circuits.

A \SI{5}{\nano\meter} thick $NbTiN$ film is deposited on a clean SOI substrate using DC reactive magnetron sputtering in a hybrid evaporator system (by AJA International). The sputtering rate and film thickness are estimated by analyzing a step structure under an atomic force microscope. The obtained film exhibits a $T_c$ of \SI{6.3}{\kelvin}, a sheet resistance of \SI{476}{\ohm\per\sq}, a residual-resistance ratio of 0.91 and an electron diffusion coefficient of \SI{0.64}{\centi\meter\squared\per\second} (see S1 in the supplementary information). Since the deposition was conducted at room temperature, the films are presumed to be amorphous. X-ray diffraction analysis revealed no peaks corresponding to $Nb_xTi_yN$ or any other phases. The SNSPD chip is patterned using a JBX-8100FS system (by JEOL) with two e-beam lithography (EBL) steps. In the first EBL step, bonding pads and electrical wiring are defined in a standard process, followed by e-beam evaporation of \SI{10}{\nano\meter}/ \SI{90}{\nano\meter} titanium-gold bilayer and lift-off. In the second EBL step, U-shaped nanowires with varying widths and lengths are defined in a \SI{50}{\nano\meter} thick e-beam resist, ZEP520A-7. This involves a high-dose exposure, rendering the resist as a negative resist \cite{oyama_study_2011, akhlaghi_waveguide_2015}. The unexposed resist is cleaned, and the pattern transfer is performed using the RIE-10NR system (by Samco) with a mixture of $SF_6$ and $O_2$, with flow rates of \SI{40}{\sccm} and \SI{10}{\sccm}, respectively, and under \SI{40}{\watt} of RF power. Finally, the remaining resist is stripped.
Each SNSPD chip features 16 nanowires positioned at the center, arranged in a 2-wire configuration with bonding pads situated at the chip edges. Fig.~\ref{fig:chip_image}(a) shows a scanning electron microscope (SEM) image of an SNSPD comprising a U-shaped nanowire with two tapered ends connected to electrical leads. The nanowires reported here are sourced from a chip containing nanowires with widths of 20, 30, 40 and \SI{50}{\nano\meter} and lengths of 10, 20, 40 and \SI{80}{\micro\meter}. The U-shaped nanowires are designed to minimize current crowding at the bends by adjusting the inner bend radii \cite{clem_geometry-dependent_2011}. Fig.~\ref{fig:chip_image}(b) presents a zoomed-in SEM image of the far end of a U-shaped \SI{50}{\nano\meter} wide $NbTiN$ nanowire with an optimized inner bend radius.\\
Fig.~\ref{fig:chip_image}(c) illustrates a schematic of the flood illumination measurement setup housed inside a cryostat combining a pulse tube cryocooler (by Sumitomo) with a $^{4}$He sorption stage (by Chase Research Cryogenics), maintaining a temperature of \SI{0.9}{\kelvin}. Light is sourced from an optical parametric oscillator (OPO), the Chromacity Auskerry model (by Chromacity) \cite{chromacity}. The OPO emits picosecond pulses at a repetition rate of \SI{100}{\mega\hertz} and spans wavelengths from \SI{1.4}{\micro\meter} to \SI{4.2}{\micro\meter} between the signal and idler outputs. In this study, the OPO output is coupled into a free space system, where neutral density filters attenuate the power to \si{\micro\watt} levels, and Spectrogon narrowband filters narrow the bandwidth to specific ranges: $2000 \pm 30$, $2328 \pm 43$, $2900 \pm 58$, and $3500 \pm 53$ \si{\nano\meter} (see S2 in the supplementary information). An illumination source at a wavelength of \SI{1550}{\nano\meter} is provided by a mode-locked laser (from KPhotonics) generating picosecond pulses at a repetition rate of \SI{50}{\mega\hertz}, followed by a programmable attenuator. A MIR fluorozirconate fiber delivers light from the free space system to the chip under test, as shown in Fig.~\ref{fig:chip_image}(c). The chip is mounted on a copper plate surrounded by a printed circuit board (PCB) equipped with 16 SMA connectors (colours are explained in the figure caption). The distance between the fiber facet and the chip surface is approximately \SI{4.3}{\milli\meter}.
Single-mode (SM) light from the MIR fiber is guaranteed, per Thorlabs specifications, between \SI{2.3}{\micro\meter} and \SI{4.1}{\micro\meter}. The emitted SM light from the end face of the fiber spreads into a Gaussian beam. For wavelengths outside the SM range, specifically at 1550 and \SI{2004}{\nano\meter}, a two-mode $LP_{11}$ Gaussian beam is considered. Using the beam divergence model, the irradiance on the chip as a function of radial distance for different wavelengths can be calculated. Fig.~\ref{fig:chip_image}(d) shows the normalized irradiance at a wavelength of \SI{2900}{\nano\meter}. We verified that all devices are adequately illuminated at the specified wavelengths.

\begin{figure}
\includegraphics[width = \columnwidth]{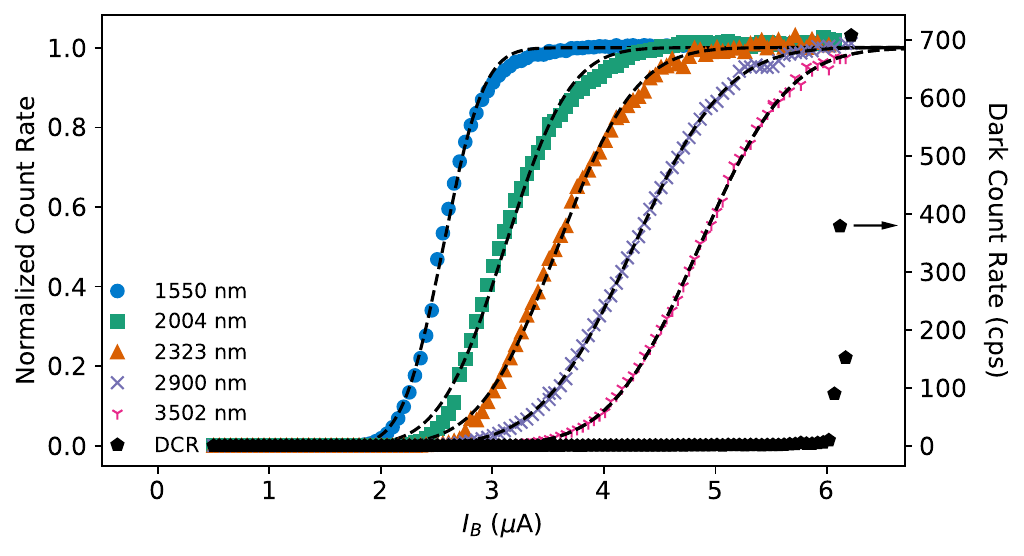}
\caption{Normalized photon count rate (i.e. IDE) from a \SI{50}{\nano\meter} wide and \SI{40}{\micro\meter} long U-shaped $NbTiN$ SNSPD at wavelengths of 1550, 2004, 2323, 2900 and \SI{3502}{\nano\meter} on the left axis, and the dark count rate on the right axis, as a function of the absolute bias current recorded at \SI{0.9}{\kelvin}. The black dashed lines represent fitted complementary error functions. The inset presents the normalized photon count rate as a function of wavelength at four different normalized bias currents ($I_{B} /I_{sw}$).}
\label{fig:CRvsIB}

    \begin{picture}(0,0)
        \put(-101.5,184.5){ \includegraphics[width=0.33\columnwidth]{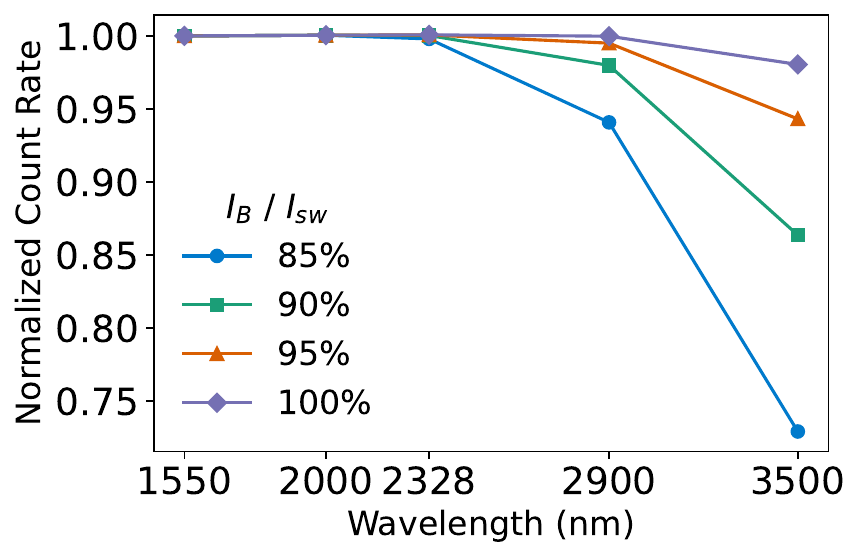}
        }
        \end{picture}
\end{figure}

The reduced cross-sectional area in these devices resulted in a limited switching current. To maximize the signal-to-noise ratio (SNR) of the readout and to prevent early latching, a modified readout circuit was adopted. The readout circuit consists of a cold off-chip LR low-pass filter in parallel with the nanowire, a room-temperature bias tee, and three low-noise amplifiers (see S3 in the supplementary information). An isolated voltage source (SIM928 by SRS), followed by a \SI{100}{\kilo\ohm} resistor, provides the biasing current. The amplified signal is recorded by a fast oscilloscope (RTO2032 by Rohde \& Schwarz) or a universal counter (53131A by Agilent).\

Fig.~\ref{fig:CRvsIB} presents normalized photon count rate (PCR) curves recorded at wavelengths 1550, 2004, 2323, 2900, and \SI{3502}{\nano\meter} on the left axis, with the dark count rate (DCR) on the right axis as a function of bias current. Coupling and absorption efficiencies were not optimized in the chip design and measurement configuration; thus, the normalized count rates here reflect the devices' IDE. All PCR curves are fitted to a simple complementary error function: $\text{PCR} = \frac{A}{2}\text{erfc}\left( \frac{I_{\text{co}} - I_B}{\Delta I_B} \right)$, where $I_{\text{co}}$ is the cut-off current defined as the current at the inflection point of the curve, $I_B$ is the bias current, and $A$ and $\Delta I_B$ are fitting parameters optimized separately for each wavelength \cite{kozorezov_fano_2017}. The fitted curves are displayed as black dashed lines. The longest saturation plateau occurs at the short wavelength of \SI{1550}{\nano\meter} and gradually shortens as the wavelength increases. According to the fits, the IDE fully saturates at all wavelengths except \SI{3502}{\nano\meter}, reaching a maximum value of 98\%. Below the saturated IDE, a given IDE value is achieved at a lower bias current for shorter wavelengths, corresponding to higher photon energy.
The inset to Fig.~\ref{fig:CRvsIB} shows the normalized photon count rate, extracted from the fitted curves, as a function of wavelength at four different normalized bias currents ($I_{B} /I_{sw}$). A cut-off in the detection efficiency of an SNSPD is observed experimentally when the bias current through the detector or the incident photon energy drops below a certain threshold \cite{semenov_spectral_2005, semenov_superconducting_2021}. A corresponding cut-off wavelength can be defined for each bias current: below this threshold, the normalized count rate remains relatively constant, while it rapidly decreases beyond it. As shown in the inset of Fig.~\ref{fig:CRvsIB}, the normalized count rate is nearly constant at a bias current close to the switching current, indicating that the cut-off wavelength is not reached. However, at a slightly lower normalized bias current, the cut-off wavelength is reached as the normalized count rate deviates from unity, resulting in a cut-off wavelength between 2900 and \SI{3502}{\nano\meter} at a normalized bias current $I_{B} /I_{sw}$ of 95\%. Further testing across more wavelengths is needed to determine the cut-off wavelength accurately. Nonetheless, it is evident that the cut-off wavelength decreases as the bias current decreases.\\

\begin{figure}
\includegraphics[width = \columnwidth]{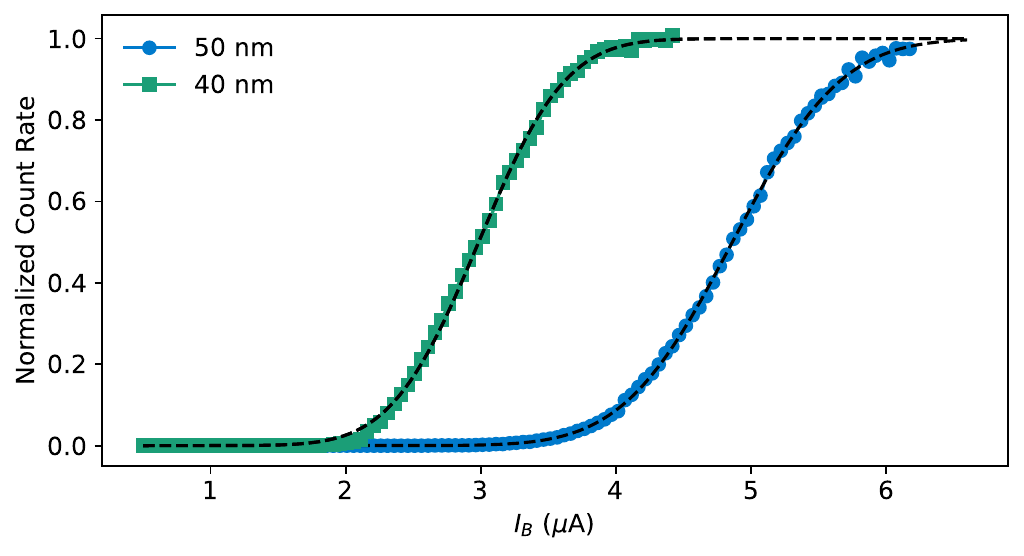}
\caption{Normalized photon count rate (i.e. IDE) of two SNSPDs with identical lengths of \SI{40}{\micro\meter}
 and widths of \SI{40}{\nano\meter} and \SI{50}{\nano\meter} at a wavelength of \SI{3502}{\nano\meter} recorded at \SI{0.9}{\kelvin}. The black dashed lines represent fitted complementary error functions.}
 \label{fig:nanowireWidth}
\end{figure}

The DCR was recorded with the MIR fiber mounted atop the chip but blocked by a metallic cap at the room-temperature optical port. It is negligible at small bias currents and increases exponentially after \SI{6}{\micro\ampere}, reaching a switching current of \SI{6.2}{\micro\ampere}. Generally, the optimal operating range for an SNSPD, in terms of bias current, is achieved at a maximized count rate (ideally, in the plateau) with minimized DCR. The extended saturation plateau observed at shorter wavelengths provides the SNSPD with a wide optimal operating range. However, this range is limited to longer wavelengths. For example, operating this detector at \SI{6}{\micro\ampere} ensures the highest IDE at all wavelengths, with a DCR of \SI{9.5}{\cps}.

A narrower nanowire with the same length was also measured for comparison. Fig.~\ref{fig:nanowireWidth} shows PCR curves of two SNSPDs with an identical length of \SI{40}{\micro\meter} and widths of 40 and \SI{50}{\nano\meter}, measured at the wavelength of \SI{3502}{\nano\meter}. The \SI{40}{\nano\meter} wide nanowire exhibits a smaller switching current but shows a saturation plateau, indicating higher sensitivity for longer wavelengths, unlike the \SI{50}{\nano\meter} wide nanowire, which reaches an IDE of 98\%. This behaviour is attributed to the reduced cross-sectional area, which increases the probability of generating a hotspot due to higher deposited energy per unit area while simultaneously reducing the total current through the device near the critical current density.

\begin{figure}
\includegraphics[width = \columnwidth]{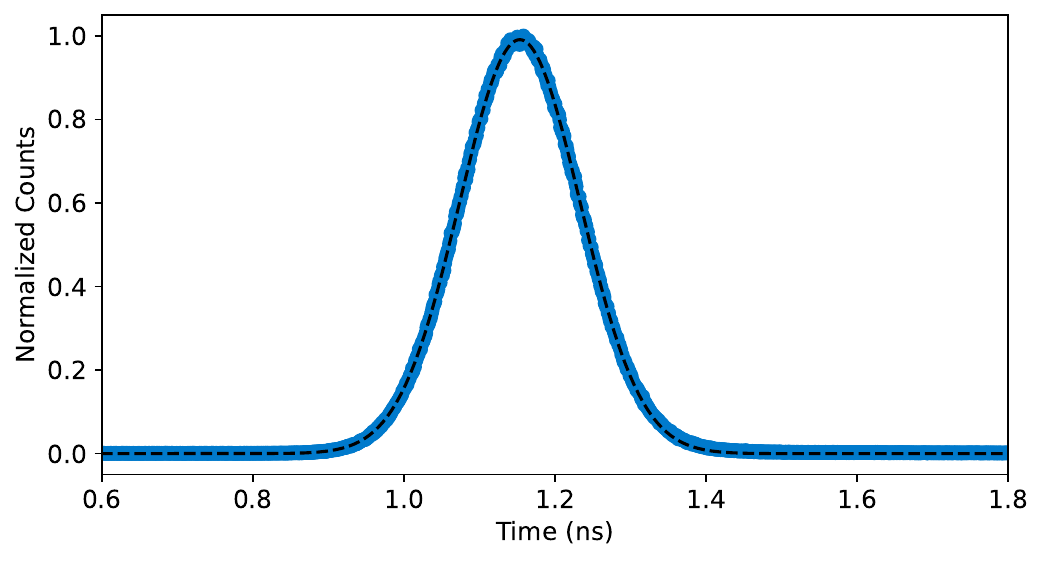}
\caption{Normalized counts histogram of a \SI{50}{\nano\meter} wide, \SI{40}{\micro\meter} long U-shaped $NbTiN$ based SNSPD at a normalized bias current $I_B/I_{sw}$ of 90\% under a wavelength of \SI{2900}{\nano\meter}. The black dashed line represents a Gaussian fit used to determine the timing jitter, yielding a FWHM of \SI{188}{\pico\second}.}
\label{fig:jitter}
\end{figure}

Timing jitter was measured at the wavelength of \SI{2900}{\nano\meter} using a time tagger (HydraHarp 400 by PicoQuant). The time tagger receives two input signals: the electrical output signal from the SNSPD and an electrical sync signal generated by a fast photodiode (DXM30AF by Thorlabs), which detects the optical input from the OPO. The jitter histogram was recorded at a normalized bias current $I_B/I_{sw}$ of 90\%, with 10,000 counts collected. Fig.~\ref{fig:jitter} displays the normalized counts histogram obtained for the \SI{50}{\nano\meter} wide SNSPD. The full-width half maximum (FWHM), extracted from a Gaussian fit, yields a jitter of \SI{188}{\pico\second}. For the application of SNSPDs in the readout of proposed spin qubits implemented in a silicon photonic platform, it is crucial that the timing jitter is shorter than the electron spin coherence time, which is \SI{2.1}{\second} at \SI{1.2}{\kelvin} \cite{morse_photonic_2017}. Our measured jitter shows a weak dependence on wavelength, ranging from 188 to \SI{196}{\pico\second}, primarily influenced by electronic jitter, affirming the suitability of our SNSPD for this demanding application. If needed, the SNSPD can be engineered for low jitter through an impedance-matched device design and implementation of a cryogenic amplifier chain \cite{zhu_superconducting_2019}.  
The recovery time of the SNSPD, which determines its maximum count rate, is defined as the duration for the voltage detection pulse to decay from 90\% to 10\% of its peak amplitude \cite{ferrari_waveguide-integrated_2018}. The recovery time for the \SI{50}{\nano\meter} wide SNSPD using our modified readout circuit is \SI{4.3}{\nano\second} (see S3 in the supplementary information). This value, smaller than those reported in the literature \cite{holzman_superconducting_2019, resta_gigahertz_2023}, suggests that achieving a count rate in the gigahertz (GHz) range is feasible through further reduction in nanowire length \cite{kerman_kinetic-inductance-limited_2006} and optimization of the readout circuit. The kinetic inductance of the nanowire, estimated based on the fall time of the SNSPD voltage detection pulse, is $L_{k} =$ \SI{145}{\nano\henry}.

\begin{figure}
\includegraphics[width = \columnwidth]{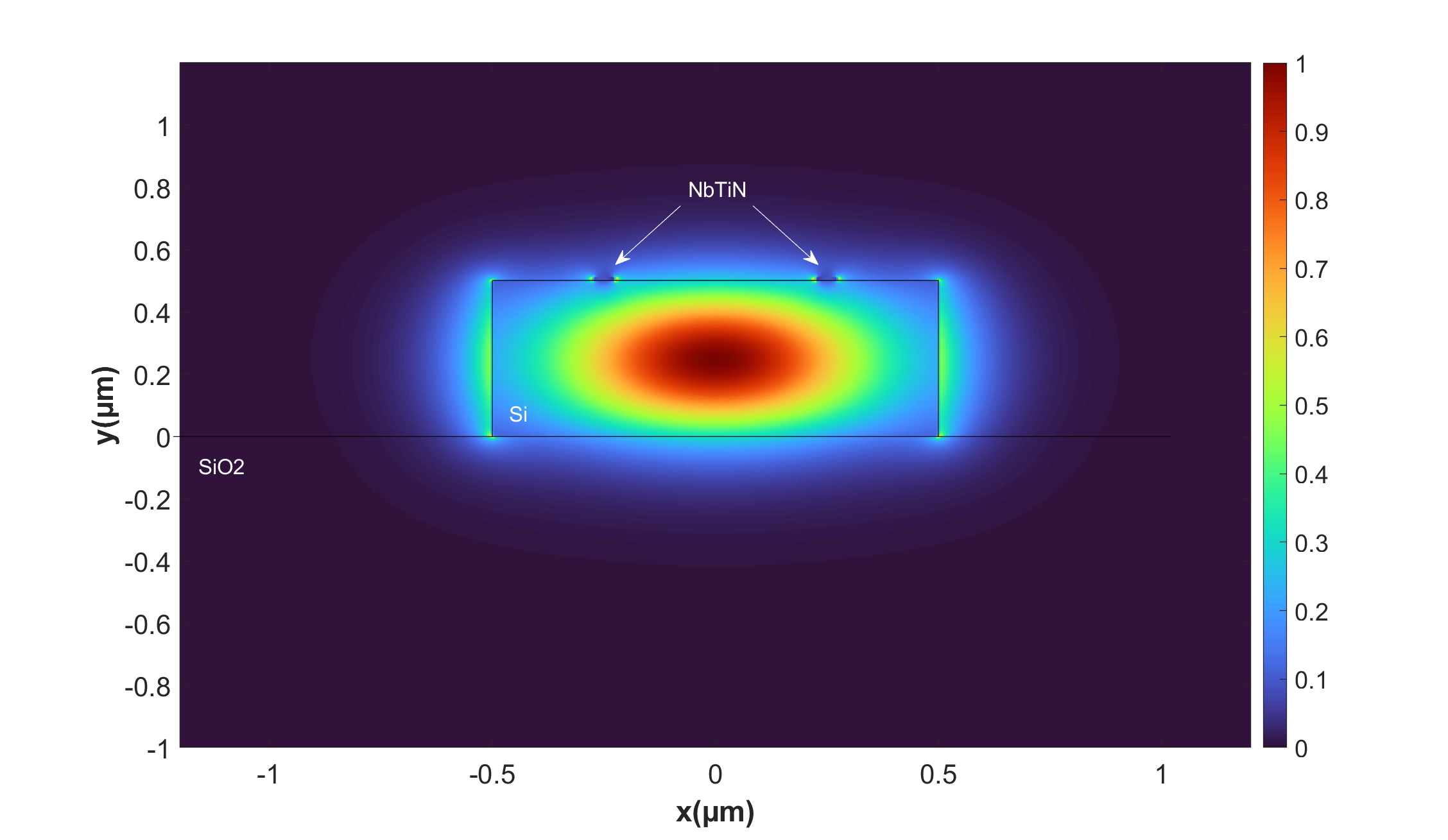}
\caption{Optical mode profile of the fundamental TE mode inside a \SI{1}{\micro\meter} wide and \SI{500}{\nano\meter} thick single-mode strip silicon waveguide on an insulating silicon dioxide substrate with two $NbTiN$ stripes on top, simulated using Lumerical FDE solver at a wavelength of \SI{2900}{\nano\meter}. The center-to-center spacing between the two stripes is \SI{500}{\nano\meter}.}
\label{fig:absorption}
\end{figure}

Reducing the cross-sectional area of the superconducting nanowire extends the spectral detection range of the SNSPD but also introduces new challenges. This reduction limits the switching current and, consequently, the bias current range. Since the SNSPD signal is directly proportional to the bias current, a decrease in bias current leads to a lower SNR, necessitating more stringent demands on the readout circuit. Moreover, it imposes stricter requirements on nanofabrication tolerances, as narrower nanowires are prone to constrictions and become highly sensitive to edge roughness. These factors might hinder achieving unity IDE, potentially affecting the yield and quality of fabricated SNSPDs. Therefore, SNSPD designs must balance achieving extended spectral sensitivity with fabrication capabilities.
Reducing the cross-sectional area of the superconducting nanowire impacts not only spectral sensitivity but also other performance metrics. Narrower nanowires are expected to exhibit lower DCR, as the breakup rate of vortex-antivortex pairs in the superconducting nanowire is proportional to its cross-sectional area \cite{engel_detection_2015}. Additionally, wider nanowires lead to a larger variation in the time duration for a vortex to cross the nanowire, thereby increasing timing jitter. Therefore, constriction-free and homogeneous narrow nanowires are favoured for improved timing jitter \cite{cheng_inhomogeneity-induced_2017}.\

Future efforts will focus on integrating the developed U-shaped SNSPDs on top of waveguides designed for operation at a wavelength of \SI{2900}{\nano\meter}. This integration will require an additional fabrication step to define a single-mode strip waveguide and other photonic components within the \SI{500}{\nano\meter} thick silicon layer in the SOI substrate. At that stage, the nanowire length should be optimized to maximize absorption efficiency while preserving fast timing response. Fig.~\ref{fig:absorption} presents the optical mode profile of the fundamental TE mode inside a strip silicon waveguide on an insulating silicon dioxide with two $NbTiN$ stripes on top. This simulation, performed using a Lumerical Finite Difference Eigenmode (FDE) solver at a wavelength of \SI{2900}{\nano\meter}, employs the refractive and extinction coefficients as a function of wavelength from a previous study \cite{banerjee_optical_2018}. The $NbTiN$ stripes are \SI{5}{\nano\meter} thick with a center-to-center spacing of \SI{500}{\nano\meter}. The evanescent light coupling from the silicon waveguide into the $NbTiN$ stripes induces an absorption loss of \SI{412}{\text{dB/cm}}, yielding an absorption efficiency of 17.3\% for a \SI{40}{\micro\meter} long nanowire. While increasing the nanowire length can improve absorption efficiency, it would limit the SNSPD's timing performance. Alternatively, embedding a short nanowire within an optical cavity could achieve unity absorption efficiency without compromising timing performance \cite{akhlaghi_waveguide_2015, vetter_cavity-enhanced_2016, munzberg_superconducting_2018}.\\

To summarize, our study demonstrates the effectiveness of reducing the cross-sectional area of $NbTiN$ nanowires to extend the spectral detection sensitivity of SNSPDs into the MIR range. Our MIR detectors are designed in a 2-wire configuration and a U-shaped geometry, the standard design for waveguide-integrated SNSPDs. Fabricated on an SOI photonic platform, our devices are optimally suited for waveguide integration, offering advantages over designs used in previous studies. The detectors exhibited near-unity IDE up to \SI{3502}{\nano\meter} under flood illumination, along with a DCR smaller than \SI{10}{\cps} at \SI{0.9}{\kelvin} and a rapid recovery time of \SI{4.3}{\nano\second}. A comparative study with a narrower SNSPD revealed enhanced sensitivity at longer wavelengths, validating the approach of reducing the nanowire cross-section.
This work establishes a robust foundation for the development of MIR waveguide-integrated SNSPDs on an SOI platform and holds significant promise for enhancing quantum communication, computing, and sensing applications in the MIR range. Future efforts will focus on demonstrating waveguide integration and achieving unity absorption efficiency through embedding the SNSPD within a photonic cavity.

\section*{Acknowledgment}
The authors thank Andreas Pfenning for introducing the two teams and initiating the collaboration.
This work was supported by the Natural Sciences and Engineering Research Council of Canada (NSERC), the Canada Foundation for Innovation (CFI), the Stewart Blusson Quantum Matter Institute (SBQMI) Advanced Nanofabrication Facility, and Innovation for Defence Excellence and Security (IDEaS). AA acknowledges financial support from the SBQMI QuEST fellowship program, NSERC CREATE in Quantum Computing Program (grant number 543245) and CMC Microsystems for the MNT award.
RHH thanks the UK Engineering and Physical Sciences Research Council (EPSRC) for support (grant awards EP/S026428/1, EP/T001011/1, EP/T00097X/1).

\section*{Data Availability}
The data that support the findings of this study are available from the corresponding author upon reasonable request.

\clearpage
\section*{References}
\addcontentsline{toc}{section}{References}
\bibliographystyle{unsrt} 
\bibliography{main}

\begin{thebibliography}{10}

\bibitem{verevkin_detection_2002}
A.~Verevkin, J.~Zhang, Roman Sobolewski, A.~Lipatov, O.~Okunev, G.~Chulkova, A.~Korneev, K.~Smirnov, G.~N. Gol’tsman, and A.~Semenov.
\newblock Detection efficiency of large-active-area {NbN} single-photon superconducting detectors in the ultraviolet to near-infrared range.
\newblock {\em Applied Physics Letters}, 80(25):4687--4689, June 2002.

\bibitem{marsili_detecting_2013}
F.~Marsili, V.~B. Verma, J.~A. Stern, S.~Harrington, A.~E. Lita, T.~Gerrits, I.~Vayshenker, B.~Baek, M.~D. Shaw, R.~P. Mirin, and S.~W. Nam.
\newblock Detecting single infrared photons with 93\% system efficiency.
\newblock {\em Nature Photonics}, 7(3):210--214, March 2013.

\bibitem{slichter_uv-sensitive_2017}
D.~H. Slichter, V.~B. Verma, D.~Leibfried, R.~P. Mirin, S.~W. Nam, and D.~J. Wineland.
\newblock {UV}-sensitive superconducting nanowire single photon detectors for integration in an ion trap.
\newblock {\em Optics Express}, 25(8):8705, April 2017.

\bibitem{morozov_superconducting_2021}
Dmitry~V. Morozov, Alessandro Casaburi, and Robert~H. Hadfield.
\newblock Superconducting photon detectors.
\newblock {\em Contemporary Physics}, 62(2):69--91, April 2021.

\bibitem{korzh_wsi_2018}
B.~Korzh, Q.-Y. Zhao, S.~Frasca, D.~Zhu, E.~Ramirez, E.~Bersin, M.~Colangelo, A.~E. Dane, A.~D. Beyer, J.~Allmaras, E.~E. Wollman, K.~K. Berggren, and M.~D. Shaw.
\newblock {WSi} superconducting nanowire single photon detector with a temporal resolution below 5 ps.
\newblock In {\em Conference on {Lasers} and {Electro}-{Optics}}, page FW3F.3, San Jose, California, 2018. OSA.

\bibitem{korzh_demonstration_2020}
Boris Korzh, Qing-Yuan Zhao, Jason~P. Allmaras, Simone Frasca, Travis~M. Autry, Eric~A. Bersin, Andrew~D. Beyer, Ryan~M. Briggs, Bruce Bumble, Marco Colangelo, Garrison~M. Crouch, Andrew~E. Dane, Thomas Gerrits, Adriana~E. Lita, Francesco Marsili, Galan Moody, Cristián Peña, Edward Ramirez, Jake~D. Rezac, Neil Sinclair, Martin~J. Stevens, Angel~E. Velasco, Varun~B. Verma, Emma~E. Wollman, Si~Xie, Di~Zhu, Paul~D. Hale, Maria Spiropulu, Kevin~L. Silverman, Richard~P. Mirin, Sae~Woo Nam, Alexander~G. Kozorezov, Matthew~D. Shaw, and Karl~K. Berggren.
\newblock Demonstration of sub-3 ps temporal resolution with a superconducting nanowire single-photon detector.
\newblock {\em Nature Photonics}, 14(4):250--255, April 2020.

\bibitem{you_superconducting_2020}
Lixing You.
\newblock Superconducting nanowire single-photon detectors for quantum information.
\newblock {\em Nanophotonics}, 9(9):2673--2692, July 2020.

\bibitem{chang_detecting_2021}
J.~Chang, J.~W.~N. Los, J.~O. Tenorio-Pearl, N.~Noordzij, R.~Gourgues, A.~Guardiani, J.~R. Zichi, S.~F. Pereira, H.~P. Urbach, V.~Zwiller, S.~N. Dorenbos, and I.~Esmaeil~Zadeh.
\newblock Detecting telecom single photons with 99.5-2.07+0.5\% system detection efficiency and high time resolution.
\newblock {\em APL Photonics}, 6(3):036114, March 2021.

\bibitem{temporao_feasibility_2008}
G.~Temporao, H.~Zibinden, S.~Tanzilli, N.~Gisin, T.~Aellen, M.~Giovannini, J.~Faist, and J.~Von Der~Weid.
\newblock Feasibility study of free-{Space} quantum key distribution in the mid-infrared.
\newblock {\em Quantum Information and Computation}, 8(1\&2):1--11, January 2008.

\bibitem{yan_silicon_2021}
Xiruo Yan, Sebastian Gitt, Becky Lin, Donald Witt, Mahssa Abdolahi, Abdelrahman Afifi, Adan Azem, Adam Darcie, Jingda Wu, Kashif Awan, Matthew Mitchell, Andreas Pfenning, Lukas Chrostowski, and Jeff~F. Young.
\newblock Silicon photonic quantum computing with spin qubits.
\newblock {\em APL Photonics}, 6(7):070901, July 2021.

\bibitem{chen_mid-infrared_2017}
Li~Chen, Dirk Schwarzer, Varun~B. Verma, Martin~J. Stevens, Francesco Marsili, Richard~P. Mirin, Sae~Woo Nam, and Alec~M. Wodtke.
\newblock Mid-infrared {Laser}-{Induced} {Fluorescence} with {Nanosecond} {Time} {Resolution} {Using} a {Superconducting} {Nanowire} {Single}-{Photon} {Detector}: {New} {Technology} for {Molecular} {Science}.
\newblock {\em Accounts of Chemical Research}, 50(6):1400--1409, June 2017.

\bibitem{dello_russo_advances_2022}
Stefano Dello~Russo, Arianna Elefante, Daniele Dequal, Deborah~Katia Pallotti, Luigi Santamaria~Amato, Fabrizio Sgobba, and Mario Siciliani De~Cumis.
\newblock Advances in {Mid}-{Infrared} {Single}-{Photon} {Detection}.
\newblock {\em Photonics}, 9(7):470, July 2022.

\bibitem{lau_superconducting_2023}
Jascha~A. Lau, Varun~B. Verma, Dirk Schwarzer, and Alec~M. Wodtke.
\newblock Superconducting single-photon detectors in the mid-infrared for physical chemistry and spectroscopy.
\newblock {\em Chemical Society Reviews}, 52(3):921--941, 2023.

\bibitem{taylor_photon_2019}
Gregor~G. Taylor, Dmitry Morozov, Nathan~R. Gemmell, Kleanthis Erotokritou, Shigehito Miki, Hirotaka Terai, and Robert~H. Hadfield.
\newblock Photon counting {LIDAR} at 2.3µm wavelength with superconducting nanowires.
\newblock {\em Optics Express}, 27(26):38147, December 2019.

\bibitem{hadfield_single-photon_2023}
Robert~H. Hadfield, Jonathan Leach, Fiona Fleming, Douglas~J. Paul, Chee~Hing Tan, Jo~Shien Ng, Robert~K. Henderson, and Gerald~S. Buller.
\newblock Single-photon detection for long-range imaging and sensing.
\newblock {\em Optica}, 10(9):1124, September 2023.

\bibitem{wollman_recent_2021}
Emma~E. Wollman, Varun~B. Verma, Alexander~B. Walter, Jeff Chiles, Boris Korzh, Jason~P. Allmaras, Yao Zhai, Adriana~E. Lita, Adam~N. McCaughan, Ekkehart Schmidt, Simone Frasca, Richard~P. Mirin, Sae~Woo Nam, and Matthew~D. Shaw.
\newblock Recent advances in superconducting nanowire single-photon detector technology for exoplanet transit spectroscopy in the mid-infrared.
\newblock {\em Journal of Astronomical Telescopes, Instruments, and Systems}, 7(01), January 2021.

\bibitem{ferrari_waveguide-integrated_2018}
Simone Ferrari, Carsten Schuck, and Wolfram Pernice.
\newblock Waveguide-integrated superconducting nanowire single-photon detectors.
\newblock {\em Nanophotonics}, 7(11):1725--1758, October 2018.

\bibitem{shankar_mid-infrared_2011}
Raji Shankar, Rick Leijssen, Irfan Bulu, and Marko Lončar.
\newblock Mid-infrared photonic crystal cavities in silicon.
\newblock {\em Optics Express}, 19(6):5579, March 2011.

\bibitem{holzman_superconducting_2019}
Itamar Holzman and Yachin Ivry.
\newblock Superconducting {Nanowires} for {Single}‐{Photon} {Detection}: {Progress}, {Challenges}, and {Opportunities}.
\newblock {\em Advanced Quantum Technologies}, 2(3-4):1800058, April 2019.

\bibitem{chen_mid-infrared_2021}
Qi~Chen, Rui Ge, Labao Zhang, Feiyan Li, Biao Zhang, Feifei Jin, Hang Han, Yue Dai, Guanglong He, Yue Fei, Xiaohan Wang, Hao Wang, Xiaoqing Jia, Qingyuan Zhao, Xuecou Tu, Lin Kang, Jian Chen, and Peiheng Wu.
\newblock Mid-infrared single photon detector with superconductor {Mo0}.{8Si0}.2 nanowire.
\newblock {\em Science Bulletin}, 66(10):965--968, May 2021.

\bibitem{taylor_low-noise_2023}
Gregor~G. Taylor, Alexander~B. Walter, Boris Korzh, Bruce Bumble, Sahil~R. Patel, Jason~P. Allmaras, Andrew~D. Beyer, Roger O’Brient, Matthew~D. Shaw, and Emma~E. Wollman.
\newblock Low-noise single-photon counting superconducting nanowire detectors at infrared wavelengths up to 29 µm.
\newblock {\em Optica}, 10(12):1672, December 2023.

\bibitem{tanner_enhanced_2010}
M.~G. Tanner, C.~M. Natarajan, V.~K. Pottapenjara, J.~A. O’Connor, R.~J. Warburton, R.~H. Hadfield, B.~Baek, S.~Nam, S.~N. Dorenbos, E.~Bermúdez Ureña, T.~Zijlstra, T.~M. Klapwijk, and V.~Zwiller.
\newblock Enhanced telecom wavelength single-photon detection with {NbTiN} superconducting nanowires on oxidized silicon.
\newblock {\em Applied Physics Letters}, 96(22):221109, May 2010.

\bibitem{miki_superconducting_2009}
Shigehito Miki, Masanori Takeda, Mikio Fujiwara, Masahide Sasaki, Akira Otomo, and Zhen Wang.
\newblock Superconducting {NbTiN} {Nanowire} {Single} {Photon} {Detectors} with {Low} {Kinetic} {Inductance}.
\newblock {\em Applied Physics Express}, 2:075002, June 2009.

\bibitem{iosad_optimization_1999}
N.N. Iosad, B.D. Jackson, T.M. Klapwijk, S.N. Polyakov, P.N. Dmitirev, and J.R. Gao.
\newblock Optimization of {RF}- and {DC}-sputtered {NbTiN} films for integration with {Nb}-based {SIS} junctions.
\newblock {\em IEEE Transactions on Appiled Superconductivity}, 9(2):1716--1719, June 1999.

\bibitem{pratap_optimization_2023}
Pratiksha Pratap, Laxmipriya Nanda, Kartik Senapati, R~P Aloysius, and Venugopal Achanta.
\newblock Optimization of the superconducting properties of {NbTiN} thin films by variation of the {N} $_{\textrm{2}}$ partial pressure during sputter deposition.
\newblock {\em Superconductor Science and Technology}, 36(8):085017, August 2023.

\bibitem{goltsman_middle-infrared_2007}
G.~Gol'tsman, A.~Korneev, M.~Tarkhov, V.~Seleznev, A.~Divochiy, O.~Minaeva, N.~Kaurova, B.~Voronov, O.~Okunev, G.~Chulkova, I.~Milostnaya, and K.~Smirnov.
\newblock Middle-infrared ultrafast superconducting single photon detector.
\newblock In {\em 2007 {Joint} 32nd {International} {Conference} on {Infrared} and {Millimeter} {Waves} and the 15th {International} {Conference} on {Terahertz} {Electronics}}, pages 115--116, Cardiff, September 2007. IEEE.

\bibitem{marsili_efficient_2012}
Francesco Marsili, Francesco Bellei, Faraz Najafi, Andrew~E. Dane, Eric~A. Dauler, Richard~J. Molnar, and Karl~K. Berggren.
\newblock Efficient {Single} {Photon} {Detection} from 500 nm to 5 µm {Wavelength}.
\newblock {\em Nano Letters}, 12(9):4799--4804, September 2012.

\bibitem{chang_efficient_2022}
Jin Chang, Johannes W.~N. Los, Ronan Gourgues, Stephan Steinhauer, S.~N. Dorenbos, Silvania~F. Pereira, H.~Paul Urbach, Val Zwiller, and Iman Esmaeil~Zadeh.
\newblock Efficient mid-infrared single-photon detection using superconducting {NbTiN} nanowires with high time resolution in a {Gifford}-{McMahon} cryocooler.
\newblock {\em Photonics Research}, 10(4):1063, April 2022.

\bibitem{taylor_mid-infrared_2022}
Gregor~G. Taylor, Ewan~N. MacKenzie, Boris Korzh, Dmitry~V. Morozov, Bruce Bumble, Andrew~D. Beyer, Jason~P. Allmaras, Matthew~D. Shaw, and Robert~H. Hadfield.
\newblock Mid-infrared timing jitter of superconducting nanowire single-photon detectors.
\newblock {\em Applied Physics Letters}, 121(21):214001, November 2022.

\bibitem{verma_single-photon_2021}
V.~B. Verma, B.~Korzh, A.~B. Walter, A.~E. Lita, R.~M. Briggs, M.~Colangelo, Y.~Zhai, E.~E. Wollman, A.~D. Beyer, J.~P. Allmaras, H.~Vora, D.~Zhu, E.~Schmidt, A.~G. Kozorezov, K.~K. Berggren, R.~P. Mirin, S.~W. Nam, and M.~D. Shaw.
\newblock Single-photon detection in the mid-infrared up to 10 µm wavelength using tungsten silicide superconducting nanowire detectors.
\newblock {\em APL Photonics}, 6(5):056101, May 2021.

\bibitem{colangelo_large-area_2022}
Marco Colangelo, Alexander~B. Walter, Boris~A. Korzh, Ekkehart Schmidt, Bruce Bumble, Adriana~E. Lita, Andrew~D. Beyer, Jason~P. Allmaras, Ryan~M. Briggs, Alexander~G. Kozorezov, Emma~E. Wollman, Matthew~D. Shaw, and Karl~K. Berggren.
\newblock Large-{Area} {Superconducting} {Nanowire} {Single}-{Photon} {Detectors} for {Operation} at {Wavelengths} up to 7.4 µm.
\newblock {\em Nano Letters}, 22(14):5667--5673, July 2022.

\bibitem{morse_photonic_2017}
Kevin~J. Morse, Rohan J.~S. Abraham, Adam DeAbreu, Camille Bowness, Timothy~S. Richards, Helge Riemann, Nikolay~V. Abrosimov, Peter Becker, Hans-Joachim Pohl, Michael L.~W. Thewalt, and Stephanie Simmons.
\newblock A photonic platform for donor spin qubits in silicon.
\newblock {\em Science Advances}, 3(7):e1700930, July 2017.

\bibitem{moody_2022_2022}
Galan Moody, Volker~J Sorger, Daniel~J Blumenthal, Paul~W Juodawlkis, William Loh, Cheryl Sorace-Agaskar, Alex~E Jones, Krishna~C Balram, Jonathan C~F Matthews, Anthony Laing, Marcelo Davanco, Lin Chang, John~E Bowers, Niels Quack, Christophe Galland, Igor Aharonovich, Martin~A Wolff, Carsten Schuck, Neil Sinclair, Marko Lončar, Tin Komljenovic, David Weld, Shayan Mookherjea, Sonia Buckley, Marina Radulaski, Stephan Reitzenstein, Benjamin Pingault, Bartholomeus Machielse, Debsuvra Mukhopadhyay, Alexey Akimov, Aleksei Zheltikov, Girish~S Agarwal, Kartik Srinivasan, Juanjuan Lu, Hong~X Tang, Wentao Jiang, Timothy~P McKenna, Amir~H Safavi-Naeini, Stephan Steinhauer, Ali~W Elshaari, Val Zwiller, Paul~S Davids, Nicholas Martinez, Michael Gehl, John Chiaverini, Karan~K Mehta, Jacquiline Romero, Navin~B Lingaraju, Andrew~M Weiner, Daniel Peace, Robert Cernansky, Mirko Lobino, Eleni Diamanti, Luis~Trigo Vidarte, and Ryan~M Camacho.
\newblock 2022 {Roadmap} on integrated quantum photonics.
\newblock {\em Journal of Physics: Photonics}, 4(1):012501, January 2022.

\bibitem{oyama_study_2011}
Tomoko~Gowa Oyama, Akihiro Oshima, Hiroki Yamamoto, Seiichi Tagawa, and Masakazu Washio.
\newblock Study on {Positive}–{Negative} {Inversion} of {Chlorinated} {Resist} {Materials}.
\newblock {\em Applied Physics Express}, 4(7):076501, June 2011.

\bibitem{akhlaghi_waveguide_2015}
Mohsen~K. Akhlaghi, Ellen Schelew, and Jeff~F. Young.
\newblock Waveguide integrated superconducting single-photon detectors implemented as near-perfect absorbers of coherent radiation.
\newblock {\em Nature Communications}, 6(1):8233, September 2015.

\bibitem{clem_geometry-dependent_2011}
John~R. Clem and Karl~K. Berggren.
\newblock Geometry-dependent critical currents in superconducting nanocircuits.
\newblock {\em Physical Review B}, 84(17):174510, November 2011.

\bibitem{chromacity}
{Chromacity Lasers}.
\newblock {Chromacity Lasers}.
\newblock \url{https://chromacitylasers.com/}.
\newblock Accessed: April 24, 2024.

\bibitem{kozorezov_fano_2017}
A.~G. Kozorezov, C.~Lambert, F.~Marsili, M.~J. Stevens, V.~B. Verma, J.~P. Allmaras, M.~D. Shaw, R.~P. Mirin, and Sae~Woo Nam.
\newblock Fano fluctuations in superconducting-nanowire single-photon detectors.
\newblock {\em Physical Review B}, 96(5):054507, August 2017.

\bibitem{semenov_spectral_2005}
A.~Semenov, A.~Engel, H.-W. Hübers, K.~Il'in, and M.~Siegel.
\newblock Spectral cut-off in the efficiency of the resistive state formation caused by absorption of a single-photon in current-carrying superconducting nano-strips.
\newblock {\em The European Physical Journal B - Condensed Matter and Complex Systems}, 47(4):495--501, October 2005.

\bibitem{semenov_superconducting_2021}
Alexej~D Semenov.
\newblock Superconducting nanostrip single-photon detectors some fundamental aspects in detection mechanism, technology and performance.
\newblock {\em Superconductor Science and Technology}, 34(5):054002, May 2021.

\bibitem{zhu_superconducting_2019}
Di~Zhu, Marco Colangelo, Boris~A. Korzh, Qing-Yuan Zhao, Simone Frasca, Andrew~E. Dane, Angel~E. Velasco, Andrew~D. Beyer, Jason~P. Allmaras, Edward Ramirez, William~J. Strickland, Daniel~F. Santavicca, Matthew~D. Shaw, and Karl~K. Berggren.
\newblock Superconducting nanowire single-photon detector with integrated impedance-matching taper.
\newblock {\em Applied Physics Letters}, 114(4):042601, January 2019.

\bibitem{resta_gigahertz_2023}
Giovanni~V. Resta, Lorenzo Stasi, Matthieu Perrenoud, Sylvain El-Khoury, Tiff Brydges, Rob Thew, Hugo Zbinden, and Félix Bussières.
\newblock Gigahertz {Detection} {Rates} and {Dynamic} {Photon}-{Number} {Resolution} with {Superconducting} {Nanowire} {Arrays}.
\newblock {\em Nano Letters}, 23(13):6018--6026, July 2023.

\bibitem{kerman_kinetic-inductance-limited_2006}
Andrew~J. Kerman, Eric~A. Dauler, William~E. Keicher, Joel K.~W. Yang, Karl~K. Berggren, G.~Gol’tsman, and B.~Voronov.
\newblock Kinetic-inductance-limited reset time of superconducting nanowire photon counters.
\newblock {\em Applied Physics Letters}, 88(11):111116, March 2006.

\bibitem{engel_detection_2015}
A~Engel, J~J Renema, K~Il’in, and A~Semenov.
\newblock Detection mechanism of superconducting nanowire single-photon detectors.
\newblock {\em Superconductor Science and Technology}, 28(11):114003, November 2015.

\bibitem{cheng_inhomogeneity-induced_2017}
Yuhao Cheng, Chao Gu, and Xiaolong Hu.
\newblock Inhomogeneity-induced timing jitter of superconducting nanowire single-photon detectors.
\newblock {\em Applied Physics Letters}, 111(6):062604, August 2017.

\bibitem{banerjee_optical_2018}
Archan Banerjee, Robert~M. Heath, Dmitry Morozov, Dilini Hemakumara, Umberto Nasti, Iain Thayne, and Robert~H. Hadfield.
\newblock Optical properties of refractory metal based thin films.
\newblock {\em Optical Materials Express}, 8(8):2072, August 2018.

\bibitem{vetter_cavity-enhanced_2016}
Andreas Vetter, Simone Ferrari, Patrik Rath, Rasoul Alaee, Oliver Kahl, Vadim Kovalyuk, Silvia Diewald, Gregory~N. Goltsman, Alexander Korneev, Carsten Rockstuhl, and Wolfram H.~P. Pernice.
\newblock Cavity-{Enhanced} and {Ultrafast} {Superconducting} {Single}-{Photon} {Detectors}.
\newblock {\em Nano Letters}, 16(11):7085--7092, November 2016.

\bibitem{munzberg_superconducting_2018}
Julian Münzberg, Andreas Vetter, Fabian Beutel, Wladick Hartmann, Simone Ferrari, Wolfram H.~P. Pernice, and Carsten Rockstuhl.
\newblock Superconducting nanowire single-photon detector implemented in a {2D} photonic crystal cavity.
\newblock {\em Optica}, 5(5):658, May 2018.

\end{thebibliography}


\begin{thebibliography}{1}

\bibitem{tinkham_introduction_2015}
Michael Tinkham.
\newblock {\em Introduction to superconductivity}.
\newblock Dover books on physics. Dover Publ, Mineola, NY, 2 ed edition, 2015.

\bibitem{vodolazov_single-photon_2017}
D.~Yu. Vodolazov.
\newblock Single-{Photon} {Detection} by a {Dirty} {Current}-{Carrying} {Superconducting} {Strip} {Based} on the {Kinetic}-{Equation} {Approach}.
\newblock {\em Physical Review Applied}, 7(3):034014, March 2017.

\bibitem{ferrari_waveguide-integrated_2018}
Simone Ferrari, Carsten Schuck, and Wolfram Pernice.
\newblock Waveguide-integrated superconducting nanowire single-photon detectors.
\newblock {\em Nanophotonics}, 7(11):1725--1758, October 2018.

\end{thebibliography}

\end{document}


\title{\Large Supplementary Information:\\ Mid-infrared characterization of NbTiN superconducting nanowire single-photon detectors on silicon-on-insulator}

\author{ \large Adan Azem,
Dmitry V. Morozov,
Daniel Kuznesof,
Ciro Bruscino,
Robert H. Hadfield,
Lukas Chrostowski,
and Jeff F. Young}

\date{ \normalsize \today}
\maketitle
\section*{ \normalsize S1- Diffusion Coefficient}
It is known that narrow (sub-micrometre width) strips formed from superconducting materials in the dirty limit, characterized by a small electron diffusion coefficient ($D$), yield SNSPDs with high single photon sensitivity \cite{tinkham_introduction_2015, vodolazov_single-photon_2017}. In the limit of a dirty superconductor, the diffusion coefficient is defined as: 
\begin{equation}
  D = - \frac{4k_{B}}{\pi e}\left( \left.\frac{dB_{c2}}{dT} \right|_{T=T_c} \right)^{-1} \label{eq1}
\end{equation}
where $e$ is the electron charge, $k_{B}$ is the Boltzmann constant, $B_{c2}$ is the second upper critical magnetic field, $T$ is the temperature, and $T_c$ is the critical temperature.

\begin{figure}[H]
  \centering
  \includegraphics[width =0.65\columnwidth]{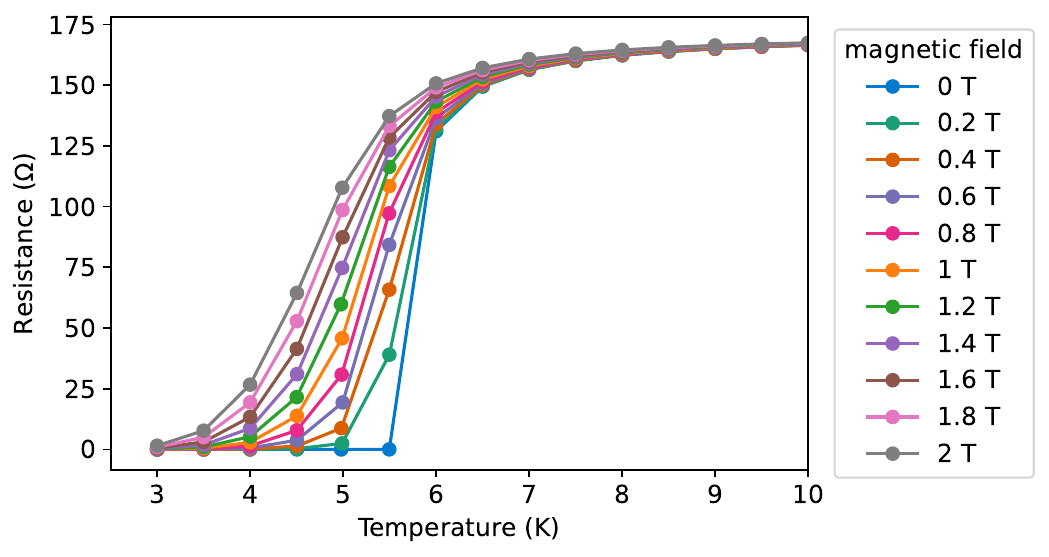}
  \caption{Resistance of a \SI{5}{\nano\meter} thick $NbTiN$ film as a function of temperature under various applied magnetic fields.}
  \label{fig:RvsT}
\end{figure}

\begin{figure}[H]
  \centering
  \includegraphics[width =0.6\columnwidth]{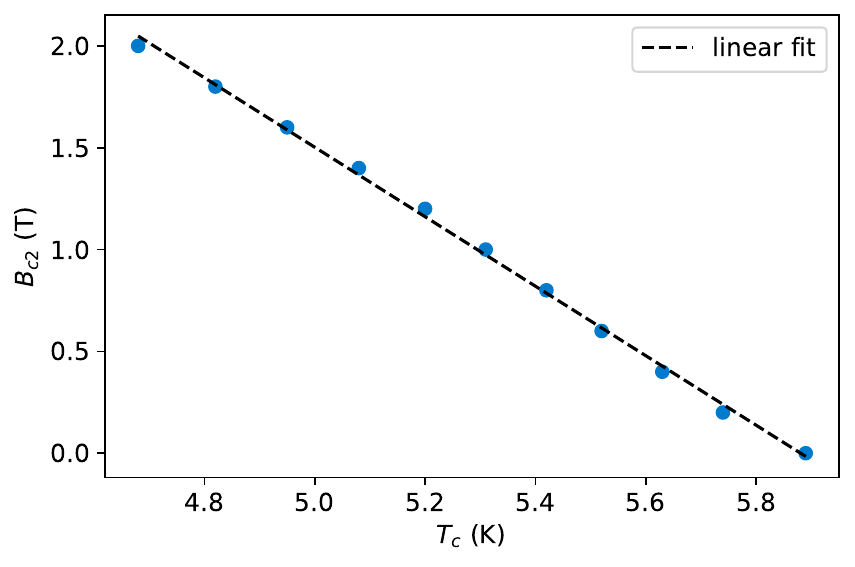}
  \caption{Upper critical magnetic field of a \SI{5}{\nano\meter} thick $NbTiN$ film as a function of critical temperature, with a linear fit.}
  \label{fig:BcvsTc}
\end{figure}

Fig.~\ref{fig:RvsT} shows the measured resistance of our $NbTiN$ thin film across varying temperatures under perpendicular magnetic fields up to \SI{2}{\tesla}. This data facilitated the extraction of the temperature dependence of the second upper critical magnetic field in the vicinity of the critical temperature, as depicted in Fig.~\ref{fig:BcvsTc}, with the critical temperature defined at the midpoint. By utilizing equation~\ref{eq1} and the slope obtained from the linear fit in Fig.~\ref{fig:BcvsTc}, we estimate the electron diffusion coefficient to be \SI{0.64}{\centi\meter\squared\per\second}.



\section*{ \large S2- Optical Spectra}
\begin{figure}[H]
  \centering
  \begin{subfigure}{0.48\textwidth}
    \includegraphics[width=\linewidth]{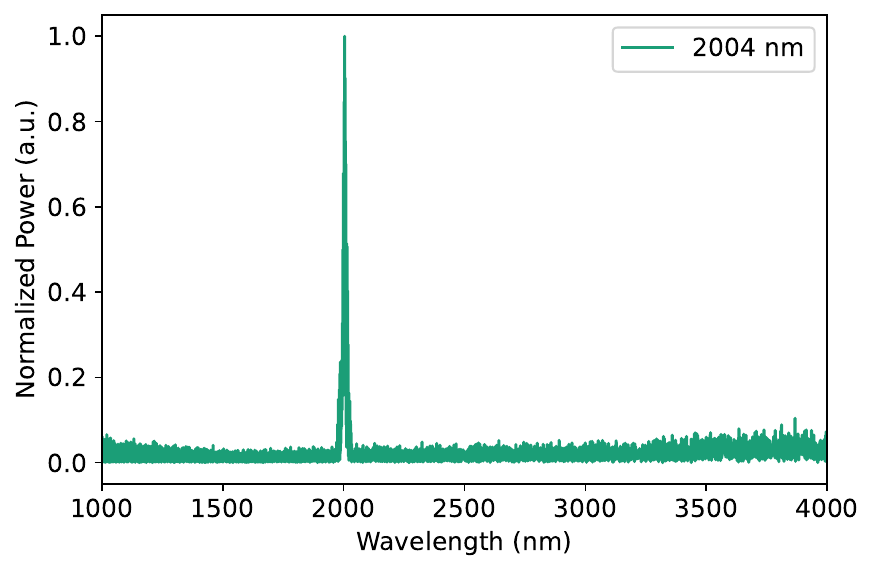}
    \caption{Signal output of OPO at $2000 \pm 30$ \si{\nano\meter}}
    \label{fig:OSA2000}
  \end{subfigure}
  \hfill
  \begin{subfigure}{0.48\textwidth}
    \includegraphics[width=\linewidth]{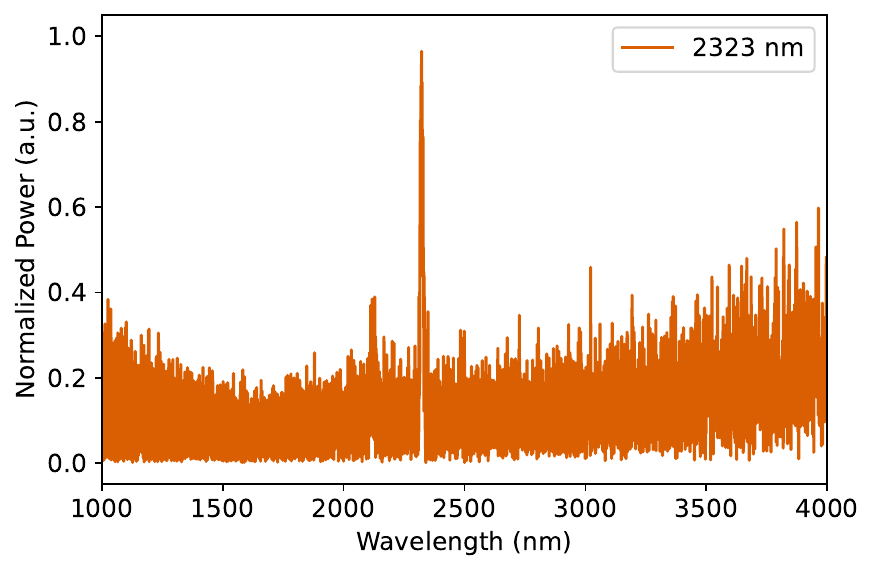}
    \caption{Idler output of OPO at $2328 \pm 43$ \si{\nano\meter}}
    \label{fig:OSA2300}
  \end{subfigure}
  \medskip
  
  \begin{subfigure}{0.48\textwidth}
    \includegraphics[width=\linewidth]{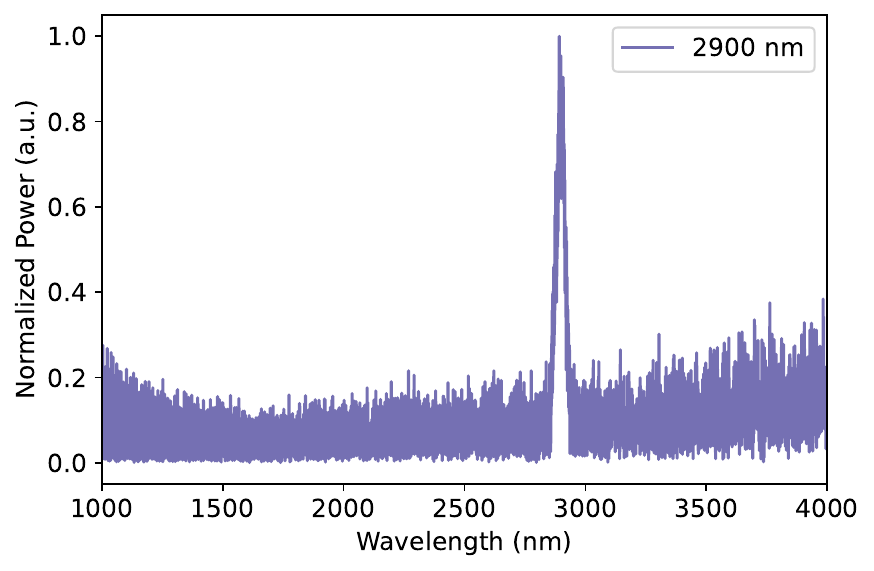}
    \caption{Idler output of OPO at $2900 \pm 58$ \si{\nano\meter}}
    \label{fig:OSA2900}
  \end{subfigure}
  \hfill
  \begin{subfigure}{0.48\textwidth}
    \includegraphics[width=\linewidth]{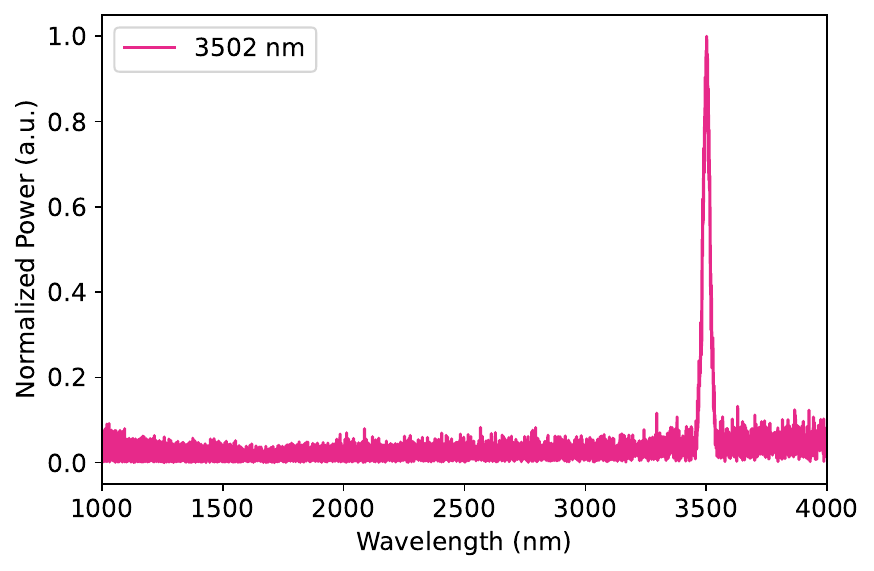}
    \caption{Idler output of OPO at $3500 \pm 53$ \si{\nano\meter}}
    \label{fig:OSA3502}
  \end{subfigure}
  \caption{Optical spectra of outputs from the OPO after filtering and attenuation, recorded by an optical spectrum analyzer.}
  \label{fig:overall}
\end{figure}

Fig.~\ref{fig:overall} displays the optical spectra of the output from the optical parametric oscillator (OPO, by Chromacity) after passing through narrowband filters and attenuators, as measured by an optical spectrum analyzer (OSA 207C by Thorlabs). The subfigures illustrate the different wavelengths obtained by the respective filters: NB-2000-030, NB-2328-043, NB-2900-058, and NB-3500-053. 
These narrowband filters effectively suppress all wavelengths outside their bandwidth, transmitting a narrow set with peak power at wavelengths of 2004, 2323, 2900, and \SI{3502}{\nano\meter}.


\section*{ \large S3- Readout Circuit and Voltage Detection Pulse}

\begin{figure}[H]
    \centering
    \begin{subfigure}{0.42\textwidth}
        \centering
        \includegraphics[width=\linewidth]{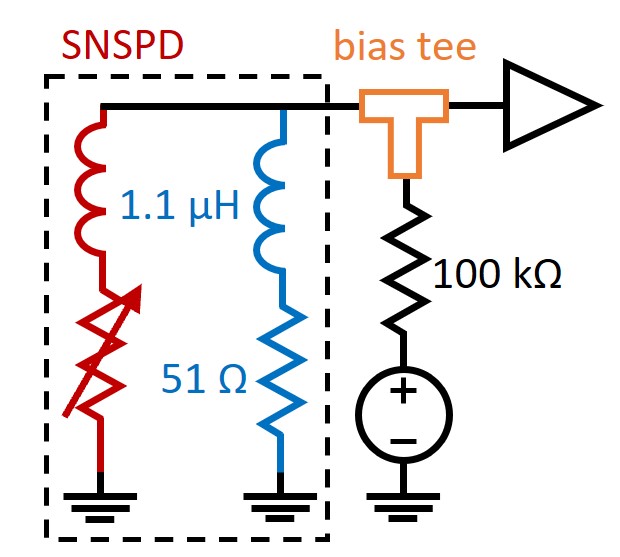}
        \caption{}
        \label{fig:readout}
    \end{subfigure}
    \hfill
    \begin{subfigure}{0.5\textwidth}
        \centering
        \includegraphics[width=\linewidth]{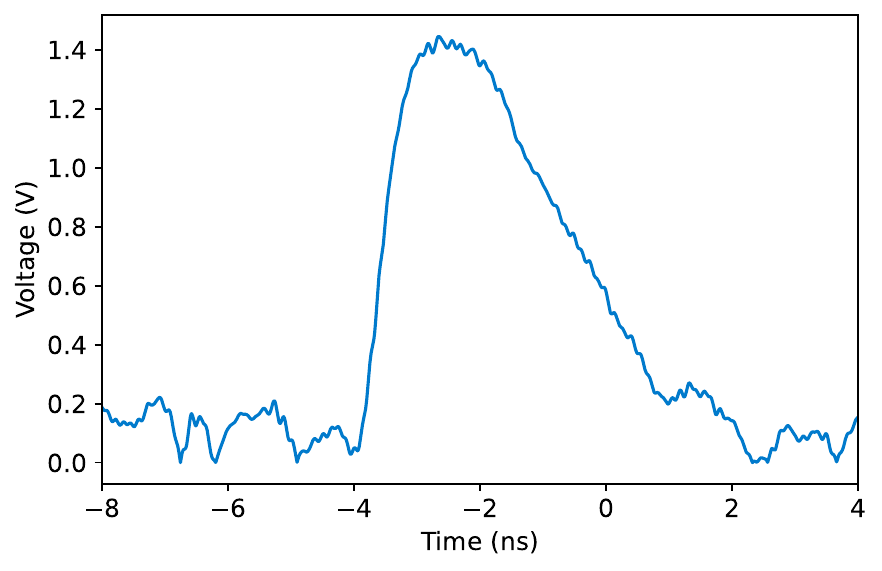}
        \caption{}
        \label{fig:voltagepulse}
    \end{subfigure}
    \caption{(a) The readout circuit: the SNSPD (red) is connected in parallel to an off-chip LR low-pass filter (blue). The dashed box specifies components mounted inside the cryostat at \SI{0.9}{\kelvin}. The room-temperature circuit includes a voltage source, biasing resistor, bias tee, and low-noise amplifiers. (b) An example of a detection voltage pulse recorded by an RTO2032 oscilloscope (\SI{3}{\giga\hertz} bandwidth) from a \SI{50}{\nano\meter} wide and \SI{40}{\micro\meter} long U-shaped $NbTiN$ SNSPD at a wavelength of \SI{1550}{\nano\meter}.}
    \label{fig:sidebyside}
\end{figure}

Fig.~\ref{fig:readout} illustrates the modified readout circuit designed to prevent latching. A superconducting nanowire, depicted as an inductor in series with a variable resistor, is connected in parallel to an off-chip cold LR low-pass filter. The room-temperature circuit includes a bias tee and three low-noise amplifiers (LNAs): one LNA-580 and two LNA-1000 from RF Bay Inc. The biasing current is provided by an isolated voltage source (SIM928 by SRS) and a \SI{100}{\kilo\ohm} resistor. The amplified signal is recorded by a fast oscilloscope (RTO2032 by Rohde \& Schwarz) or a universal counter (53131A by Agilent).\

Fig.~\ref{fig:voltagepulse} shows a voltage detection pulse captured by an oscilloscope. The trace demonstrates a rapid rise followed by exponential decay. The rise time is constrained by the readout circuitry, while the recovery time, defined as the duration for the pulse to decay from 90\% to 10\% of its peak amplitude \cite{ferrari_waveguide-integrated_2018}, is determined from an exponential decay fit and yields \SI{4.3}{\nano\second}.

\addcontentsline{toc}{section}{References}
\bibliographystyle{unsrt} 
\bibliography{main_SI}